\documentclass[12pt]{article}

\usepackage{a4wide}
\usepackage{cite}
\usepackage{amssymb}
\usepackage{amsmath}
\usepackage{epsfig}
\usepackage{eucal}

\newlength{\horizMargin}
\newcommand{\almu}{\alpha_{\mu}}         
\newcommand{\as}{\alpha_\mathrm{s}}      
\newcommand{\asb}{\bar{\alpha}_\mathrm{s}}
\newcommand{\chie}{\chi_{\e}}            
\newcommand{\dif}{\mathrm{d}}            
\newcommand{\e}{\varepsilon}

\newcommand{\esp}[1]{\mathrm{e}^{#1}}    
\newcommand{\ff}{\mathcal{N}}            
\newcommand{\G}{\CMcal{G}}               
\newcommand{\gammaL}{\gamma_{\mathrm{L}}}
\newcommand{\GeV}{\;\mathrm{GeV}}
\newcommand{\half}{{\textstyle\frac{1}{2}}}
\newcommand{\II}{\mathbb{I}}             
\renewcommand{\Im}{\;\CMcal{I}m}         
\newcommand{\Kernel}{\mathcal{K}_{\e}}   
\newcommand{\kt}{\boldsymbol{k}}          
\newcommand{\lt}{\boldsymbol{l}}          
\newcommand{\MSbar}{\overline{\mathrm{MS}}}
\newcommand{\NL}{{\mathrm{NL}x}}
\newcommand{\NLLB}{\mathrm{NLL_{B}}}
\newcommand{\NLLBp}{\mathrm{NLL_{B'}}}
\newcommand{\om}{\omega}
\newcommand{\ord}[1]{\CMcal{O}\left(#1\right)}
\newcommand{\qKernel}[1][p]{H_{\e}^{(#1)}}
\newcommand{\qRat}{q_{\mathrm{rat}}^{(\MSbar)}}
\newcommand{\qRes}{\mathcal{H}}          
\newcommand{\qResR}{\qRes^{\mathrm{rat}}}
\newcommand{\qResT}{\qRes^{\mathrm{tran}}}
\newcommand{\qcf}[1][p]{\qRes^{(#1)}}    
\newcommand{\rk}{\mathfrak{R}}           
\renewcommand{\sp}{\mathrm{sp}}          
\newcommand{\ugd}{{\CMcal F}}            
\newcommand{\ui}{\mathrm{i}}             

\begin{document}


\titlepage

\begin{flushright}
  BNL-NT-06/4\\  DFF 432/01/06\\  LPTHE-06-01\\  hep-ph/0601200 \\
  January 2006
\end{flushright}

\vspace*{1in}

\begin{center}
  {\Large \bf
  Minimal Subtraction vs.\ Physical Factorisation Schemes\\[3mm]
  in Small-$\boldsymbol{x}$ QCD}
  
  \vspace*{0.4in}
  
  M.~Ciafaloni$^{(a)}$,
  D.~Colferai$^{(a)}$,
  G.P.~Salam$^{(b)}$
  and A.M.~Sta\'sto$^{(c)}$ \\
  
  {\small
  \vspace*{0.5cm}
  $^{(a)}$ {\it  Dipartimento di Fisica, Universit\`a di Firenze,
   50019 Sesto Fiorentino (FI), Italy}; \\
  \vskip 2mm
  {\it  INFN Sezione di Firenze,  50019 Sesto Fiorentino (FI), Italy}\\
  \vskip 2mm
  $^{(b)}$ {\it LPTHE, Universit\'e Pierre et Marie Curie -- Paris 6,
    Universit\'e Denis Diderot -- Paris 7, CNRS UMR 7589,
  75252 Paris 75005, France}\\
  \vskip 2mm
  $^{(c)}$ {\it Physics Department, Brookhaven National Laboratory, Upton, NY 11973, USA};\\
  and\\
  {\it H.~Niewodnicza\'nski Institute of Nuclear Physics, Krak\'ow, Poland}\\
  \vskip 2mm}
\end{center}


\bigskip

\begin{abstract}
  \noindent
  We investigate the relationship of ``physical'' parton densities
  defined by $\kt$-factorisation, to those in the minimal subtraction
  scheme, by comparing their small-$x$ behaviour. We first summarize
  recent results on the above scheme change derived from the BFKL
  equation at NL$x$ level, and we then propose a simple extension to
  the renormalisation-group improved (RGI) equation. In this way we
  are able the examine the difference between resummed gluon
  distributions in the $Q_0$ and $\MSbar$ schemes and also to show
  $\MSbar$ scheme resummed results for $P_{gg}$ and approximate
  ones for $P_{qg}$. We find that, due to
  the stability of the RGI approach, small-$x$ resummation effects are
  not much affected by the scheme-change in the gluon channel, while
  they are relatively more sensitive for the quark--gluon mixing.
\end{abstract}

\newpage


Predictions of perturbative Quantum Chromodynamics for DGLAP~\cite{DGLAP} evolution
kernels in hard processes have been substantially improved in the past
few years, both by higher order calculations for any Bjorken
$x$~\cite{NNLODGLAP} and by resummation methods in the small-$x$
region~\cite{Salam1998,CC,CCS1,CCSSkernel,SCHMIDT,FRSV,THORNE,%
  ABF2000,ABF2001,ABF2003,ABF2005,BF2006}.  However, higher order
splitting functions are factorisation-scheme dependent: while the NNLO
results and standard parton densities~\cite{MRST,CTEQ} are obtained in
the minimal subtraction ($\MSbar$) scheme, the resummed ones are in
the so-called $Q_0$-scheme, in which the gluon density is defined by
$\kt$-factorisation of a physical process. Therefore, in order to
compare theoretical results, or to exploit the small-$x$ results in
the analysis of data, we need a precise understanding of the
relationship of physical schemes based on $\kt$-factorisation and of
minimal subtraction ones, with sufficient accuracy.

The starting point in this direction is the work of Catani, Hautmann
and one of us (M.C.)~\cite{CaCiHa93,CaHa94}, who calculated the
leading-$\log x$ (LL$x$) coefficient function $R$ of the gluon density
in the (dimensional) $Q_0$-scheme%
\footnote{The label $Q_0$ referred originally~\cite{Q0scheme} to the
  fact that the initial gluon, defined by $\kt$-factorisation, was set
  off-mass-shell ($\kt^2 = Q_0^2$) in order to cutoff the infrared
  singularities. It turns out~\cite{dimensional}, however, that the
  effective anomalous dimension at scale $\kt^2 \gg Q_0^2$ is
  independent of the cut-off procedure, whether of dimensional type or
  of off-mass-shell one.}  versus the minimal subtraction one, namely
\begin{equation}\label{schemeRel}
  g^{(Q_0)}(t,\om) = R\Big(\gammaL\big(\frac{\asb(t)}{\om}\big)\Big)
  g^{(\MSbar)}(t,\om) \;, \quad t \equiv \log\frac{\kt^2}{\mu^2}
  \;, \quad \asb \equiv \as \frac{N_c}{\pi}
\end{equation}
where $\om$ is the moment index conjugated to $x$, the $\MSbar$ density
\begin{equation}\label{gMSbar}
  g^{(\MSbar)}(t,\om) = \exp\left[\frac1{\e}\int_0^{\asb(t)/\om}
 \frac{\dif a}{a} \;\gammaL(a) \right]
\end{equation}
factorises a string of minimal-subtraction $1/\e$ poles starting from
an on-shell massless gluon, $\gammaL(\asb/\om)$ is the LL$x$
BFKL~\cite{BFKL} anomalous dimension, and the explicit form of $R$
will be given shortly.

The purpose of the present Letter is to show how to generalise the
relation~(\ref{schemeRel}) to possibly resummed subleading-log levels
and to quarks. We first summarize the essentials of such
generalisation at next-to-leading $\log(x)$ (NL$x$) level, following
the detailed analysis of the BFKL equation in $4+2\e$ dimensions of
two of us~\cite{dimensional}. We then propose a simple extension of
the method to the renormalisation group improved (RGI)
approach~\cite{CC,CCSSkernel}. On this basis, we show the effect of a
$Q_0$ to $\MSbar$ scheme change on a toy resummed gluon distribution
and we calculate a full $\MSbar$ small-$x$ resummed $P_{gg}$ evolution
kernel as well as a small-$x$ resummed $P_{qg}$ evolution kernel
in an approximation to the $\MSbar$ scheme.

\section{Scheme change to the $\boldsymbol\MSbar$ gluon at NL$\boldsymbol x$ level}

Let us first summarize the results of~\cite{dimensional} for the gluon
channel only ($N_f = 0$). The starting point is the BFKL
equation~\cite{BFKL} with NL$x$ corrections~\cite{FaLi98,CaCi98}
continued to $4+2\e$ dimensions, as described in more detail
in~\cite{dimensional}. In particular, we consider running coupling
evolution at the level of the one-loop $\beta$-function
\begin{equation}\label{beta}
  \beta(\as,\e) = \e \as - b \as^2 \;, \quad b = \frac{11 N_c}{12 \pi}
\end{equation}
so that
\begin{equation}\label{alfa}
  \frac1{\as(t)} - \frac{b}{\e} = \esp{-\e t}
  \left(\frac1{\almu}-\frac{b}{\e}\right) \;, \quad
 \as(t) = \frac{\almu \esp{\e t}}{1+b\almu\frac{\esp{\e t}-1}{\e}} \;,
\end{equation}
where $\almu\equiv(g\mu^{\e})^2\esp{-\e \psi(1)}/(4\pi)^{1+\e}$ is normalised
according to the $\MSbar$ scheme.

Note first that, the ultraviolet (UV) fixed point of Eq.~(\ref{beta})
at $\as = \e/b$ separates the evolution~(\ref{alfa}) into two distinct
regimes, according to whether {\it (i)} $\almu < \e/b$ or {\it (ii)}
$\almu > \e/b$. In the regime {\it (i)} $\as(t)$ runs monotonically
from $\as = 0$ to $\as = \e/b$ for $-\infty < t < +\infty$ --- and is
thus infrared (IR) free and bounded ---, while in the regime {\it
  (ii)} $\as(t)$ starting from $\e/b$ in the UV limit, goes through
the Landau pole at $t_\Lambda = \log(1-\e/b\almu)/\e < 0$, and reaches
$\as = 0$ from below in the IR limit.

The main result of~\cite{dimensional} is a factorisation formula for
the BFKL gluon density in $4+2\e$ dimensions. If the gluon is
initially on-shell, and the ensuing IR singularities are regulated by
$\e>0$ in the (unphysical) running coupling regime {\it (i)} mentioned
above, then the gluon density at scale $\kt^2 = \mu^2 \esp{t}$
factorises in the form
\begin{equation}\label{factor}
  g_{\e}^{(Q_0)}(t,\om) = \ff_{\e}\big(\as(t),\om\big) \exp\int_{-\infty}^t
  \dif\tau \; \bar{\gamma}\big(\as(\tau),\om;\e\big) \;,
\end{equation}
where $\bar{\gamma}$ is the saddle point value of the anomalous
dimension variable $\gamma$ conjugated to $t$ and the factor
$\ff_{\e}$ --- which is perturbative in $\as(t)$ and $\e$ --- is due
to fluctuations around the saddle point.  The expression of
$\bar{\gamma}$ for $\e > 0$ is determined by the analogue of the BFKL
eigenvalue function, namely at NL$x$ level by the equation
\begin{equation}\label{gammabar}
 \asb(t) \left[ \chie^{(0)}(\bar{\gamma})
 + \om \frac{\chie^{(1)}(\bar{\gamma})}{\chie^{(0)}(\bar{\gamma}-\e)}
 \right] = \om \;,
\end{equation}
where, by definition,
\begin{align}
 \Kernel^{(0)}\;(\kt^2)^{\gamma-1-\e}  &= \chie^{(0)}(\gamma)\, (\kt^2)^{\gamma-1} \;,
\label{chi0} \\
 \Kernel^{(1)}\;(\kt^2)^{\gamma-1-2\e} &= \chie^{(1)}(\gamma)\, (\kt^2)^{\gamma-1} \;,
\label{chi1}
\end{align}
and the detailed form of the kernels is found in
Refs.~\cite{FaLi98,CaCi98} on the basis of Refs.~\cite{RGvert,QQvert}.
The result~(\ref{factor}) is proven in~\cite{dimensional} from the
Fourier representation of the solutions of the NL$x$ equation by using
a saddle-point method in the limit of small $\e=\ord{b\as}$, which
singles out $\bar{\gamma}$ as in Eq.~(\ref{gammabar}).

Let us now make the key observation that --- due to the infinite IR
evolution down to $\as(-\infty) = 0$ --- the exponent in
Eq.~(\ref{factor}) develops $1/\e$ singularities according to the
identity
\begin{equation}\label{identity}
 \int_{-\infty}^t \dif\tau\; \bar{\gamma}\big(\as(\tau),\om;\e\big)
  = \int_0^{\as(t)} \frac{\dif\alpha}{\alpha(\e-b\alpha)} \;
 \bar{\gamma}(\alpha,\om;\e) \;,
\end{equation}
which produces single and higher order $\e$-poles in the formal
$b\alpha/\e$ expansion of the denominator. However, such singularities
are not yet in minimal subtraction form, because we can expand in the
$\e$ variable the leading and NL parts of $\bar{\gamma}$, as follows:
\begin{align}
 \bar{\gamma}(\as,\om;\e) &= \gamma^{(0)}\Big(\frac{\asb}{\om},\e\Big)
 + \as \gamma^{(1)}\Big(\frac{\asb}{\om},\e\Big)
\label{gammaExp} \\
 &= \gamma_0^{(0)}\Big(\frac{\asb}{\om}\Big)
 + \as \gamma_0^{(1)}\Big(\frac{\asb}{\om}\Big)
 + \e \gamma_1^{(0)}\Big(\frac{\asb}{\om}\Big)
 + \e \left[ \as \gamma_1^{(1)}\Big(\frac{\asb}{\om}\Big)
             +\e \gamma_2^{(0)}\Big(\frac{\asb}{\om}\Big)
 \right] + \cdots \;.
\nonumber
\end{align}
While the $\e = 0$ part is already in minimal subtraction form, the terms
$\ord{\e}$ and higher need to be expanded in the variable $\e - b\alpha$ in
order to cancel the series of $\e$-poles generated by the denominator:
\begin{subequations}\label{denomExp}
\begin{align}
  \frac{\e}{\e-b\alpha} &= \frac{b\alpha}{\e-b\alpha} + 1 \;, \\
  \frac{\e^2}{\e-b\alpha} &= \frac{b^2\alpha^2}{\e-b\alpha} + b\alpha + \e  \;,
\end{align}
\end{subequations}
and similarly for the higher order terms in $\e$.
Therefore, by replacing Eqs.~(\ref{gammaExp}) and (\ref{denomExp})
into Eq.~(\ref{factor}), we are able to factor out the minimal
subtraction density in the form
\begin{align}
  g_{\e}^{(Q_0)}(t,\om) &= R_{\e}\big(\as(t),\om\big) \exp\int_0^{\as(t)}
  \frac{\dif\alpha}{\alpha(\e-b\alpha)} \; \gamma^{(\MSbar)}(\alpha,\om)
 \nonumber \\
  &= R_{\e}\big(\as(t),\om\big)  g_{\e}^{(\MSbar)}(t,\om) \;,
 \label{Rg}
\end{align}
where now the $\e$-independent $\MSbar$ anomalous dimension is
\begin{align}
  \gamma^{(\MSbar)}(\as,\om) &= \bar{\gamma}(\as,\om;b\as)
 \label{gammaMSbar} \\
  &= \gamma_0^{(0)}\Big(\frac{\asb}{\om}\Big)
  + \as \gamma_0^{(1)}\Big(\frac{\asb}{\om}\Big)
  + b\as \left[ \gamma_1^{(0)}\Big(\frac{\asb}{\om}\Big)
             + \as \gamma_1^{(1)}\Big(\frac{\asb}{\om}\Big)
             + b\as \gamma_2^{(0)}\Big(\frac{\asb}{\om}\Big)
  \right] \, ,
 \nonumber
\end{align}
and contains some NNL$x$ terms related to the $\e$-dependent ones in
square brackets in Eq.~(\ref{gammaExp}).

Correspondingly, the coefficient $R_{\e}$ in Eq.~(\ref{Rg}) has a {\em
  finite} $\e = 0$ limit, at fixed values of $\almu$ and $\as(t) =
\almu/(1+b\almu t)$. Therefore, we are able to reach the physical
UV-free regime {\it (ii)}, and we obtain
\begin{equation}\label{rk}
 \frac{R_0\big(\as(t),\om\big)}{\ff_0\big(\as(t),\om\big)} \equiv
 \rk\big(\as(t),\om\big) = \exp\int_0^{\asb(t)} \frac{\dif\alpha}{\alpha} \;
 \left[ \gamma_1^{(0)}\Big(\frac{\alpha}{\om}\Big) + \left(
   \alpha \gamma_1^{(1)}\Big(\frac{\alpha}{\om}\Big) + b\alpha
   \gamma_2^{(0)}\Big(\frac{\alpha}{\om}\Big) \right) \right] \;,
\end{equation}
which is the result for the $R$ factor at NL$x$ level we were looking
for.

A few remarks are in order. Firstly, the expansion coefficients
$\gamma_1^{(0)}$ and $\gamma_2^{(0)}$ are simply obtained from the
known form of the $\e$-dependence of $\chie^{(0)}(\gamma) \equiv
\chi_0(\gamma) + \e\chi_1(\gamma)+\e^2\chi_2(\gamma)+\ord{\e^3}$ in
Eq.~(\ref{chi0}), while $\gamma_1^{(1)}$ is not explicitly known,
because the $\e$-dependence of $\chie^{(1)}(\gamma)$ in
Eq.~(\ref{chi1}) has yet to be extracted from the
literature~\cite{RGvert,QQvert}. We quote the results
\begin{align}
  \frac{\asb}{\om}\chi_0\big(\gamma_0^{(0)}\big) &= 1
 \label{gamma0} \\
  \gamma_1^{(0)} &= \left. -\frac{\chi_1(\gamma)}{\chi_0'(\gamma)}
  \right|_{\gamma = \gamma_0^{(0)}(\frac{\asb}{\om})}
 \label{gamma1} \\
  \gamma_2^{(0)} &= \left. -\frac{\chi_2(\gamma)}{\chi_0'(\gamma)}
  + \frac{\chi_1(\gamma) \chi_1'(\gamma)}{\chi_0'{}^2(\gamma)}
  -\frac12\frac{\chi_1^2(\gamma) \chi_0''(\gamma)}
  {\chi_0'{}^3(\gamma)} \right|_{\gamma = \gamma_0^{(0)}(\frac{\asb}{\om})} \;.
 \label{gamma2}
\end{align}
In particular $\gamma_1^{(0)}$, together with the LL$x$ form of
\begin{equation}\label{ff0}
 \ff_0 = \frac1{\gammaL \sqrt{-\chi'{}^{(0)}(\gammaL)}} \;,
 \quad \gammaL \equiv \gamma_0^{(0)} \; ,
\end{equation}
yields the result of Ref.~\cite{CaCiHa93,CaHa94}
\begin{subequations}
\label{R0}
\begin{align}
  R_0\big(\as(t),\om\big) &=
  R\Big(\gammaL\big(\frac{\asb(t)}{\om}\big)\Big) =
  \frac1{\gammaL
    \sqrt{-\chi'{}^{(0)}(\gammaL)}} \exp\int_0^{\asb(t)}
  \frac{\dif\alpha}{\alpha}\; \left[
    \gamma_1^{(0)}\Big(\frac{\alpha}{\om}\Big) + \NL \right]
 \\ 
\label{R0b}
&= \left\{ \frac{\Gamma(1-\gammaL)\chi_0(\gammaL)}
  {\Gamma(1+\gammaL)[-\gammaL\chi_0'(\gammaL)]} \right\}^{1/2}
\exp\left\{\gammaL\,\psi(1)+\int_0^{\gammaL}\dif\gamma'\;
  \frac{\psi'(1)-\psi'(1-\gamma')}{\chi_0(\gamma')}\right\} \;,
\end{align}
\end{subequations}
while $\gamma_2^{(0)}$ and $\gamma_1^{(1)}$ provide the new NL$x$ contribution
to $R$ of~\cite{dimensional}.

Secondly, the anomalous dimension in the $Q_0$-scheme takes
contributions from
$\ff_0$ only, namely%
\footnote{The normalisation factor $\ff_0$ takes NL$x$ corrections
  too, which however coincide~\cite{dimensional} with those obtained
  by the known fluctuation expansion at $\e = 0$.}
\begin{equation}\label{gammaQ0}
  \gamma^{(Q_0)}(\as,\om) = \gamma_0^{(0)}\Big(\frac{\asb}{\om}\Big)
  + \as \gamma_0^{(1)}\Big(\frac{\asb}{\om}\Big) - b\as^2
  \frac{\partial\log \ff_0(\as,\om)}{\partial\as} \;,
\end{equation}
and is therefore independent of the kernel properties for $\e > 0$. On
the other hand, by Eqs.~(\ref{gammaMSbar}) and (\ref{rk}),
$\gamma^{(\MSbar)}$ is related to $\rk$ by the expression
\begin{equation}\label{gammaMS}
 \gamma^{(\MSbar)}(\as,\om) = \gamma_0^{(0)}\Big(\frac{\asb}{\om}\Big)
  + \as \gamma_0^{(1)}\Big(\frac{\asb}{\om}\Big) + b\as^2
  \frac{\partial\log \rk(\as,\om)}{\partial\as} \;,
\end{equation}
whose origin is tied up to the identity~(\ref{denomExp}). Indeed, we
have separated terms of order $\e$ or $\e^2$ into minimal subtraction
and coefficient contributions: therefore, their $t$-evolution should
cancel out in the $\e = 0$ limit, which is the content of
Eq.~(\ref{gammaMS}).

Using Eqs.~(\ref{gammaQ0}) and (\ref{gammaMS}), the well known
relations~\cite{CaCiHa93} for NL$x$ anomalous dimensions can be
extended to subleading levels, as generated by the
$\e$-expansion. Thus the difference
\begin{equation}\label{diffAnomDim}
 \gamma^{(\MSbar)} -\gamma^{(Q_0)} = b \as \left[ \gamma_1^{(0)}
 + \as \gamma_1^{(1)} + b \as \gamma_2^{(0)}
 + \frac{\partial\log \ff_0}{\partial\log\as} \right] \;.
\end{equation}
is computed up to NNL$x$ level as outlined above, even if the
dynamical $\e = 0$ NNL$x$ contributions to the $\gamma$'s are not
investigated here.

We conclude that, while the anomalous dimension in the $Q_0$-scheme
(which is roughly a ``maximal'' subtraction one) only depends on the
$\e = 0$ properties of the BFKL evolution, the $\MSbar$ coefficient
and anomalous dimension both depend on higher orders in the
$\e$-expansion of the kernel eigenvalue, which generate subleading
contributions. The result in Eq.~(\ref{diffAnomDim}) of
~\cite{dimensional} directly provides the scheme change for the gluon
anomalous dimension at NNL$x$ level.

\section{Resummed results for the $\boldsymbol\MSbar$ gluon splitting function}

A problem exists concerning the magnitude of the scheme change
summarized above in the small-$x$ region. In fact, the explicit form
of the coefficients $\ff$, $\gamma_1^{(0)}$, $\gamma_1^{(1)}$,
$\gamma_2^{(0)}$ exhibit leading Pomeron singularities of
increasing weight, indicating that a small-$x$ resummation is in
principle required for the scheme-change too.  As a consequence, any
resummed evolution model should provide, in principle, information on
the corresponding $\e$-dependence for a rigorous relation to the
$\MSbar$ factorisation scheme, a task which appears to be practically
impossible.

In order to circumvent this difficulty, we remark that $R$ in Eq.~(\ref{R0b}) is
directly expressed as a function of the variable $\gamma$, and that the leading
Pomeron singularity occurs because of the saddle point identification
$\gamma = \gammaL\big(\asb(t)/\om)\big)$. It is then conceivable that such a
singularity will be replaced by a much softer one if the effective anomalous
dimension variable becomes $\gamma = \gamma_{\mathrm{res}}\big(\asb(t)/\om)\big)$
at resummed level.  A replacement similar to this one was used in anomalous
dimension space in the study of the scheme-change of~\cite{ABF2000}. In our
framework, we are able to ensure in general that $\gamma_{\mathrm{res}}$ is the
relevant variable by assuming that the $Q_0 \to \MSbar$ normalisation change
occurs in a $\kt$-factorised form, i.e., by taking the ``ansatz''
\begin{equation}\label{ktAnsatz}
 g_{\om}^{(\MSbar)}(t) = \int\frac{\dif\gamma}{2\pi\ui}\; \esp{\gamma t}
 \frac1{\gamma\,\tilde{R}(\gamma,\om)} f_{\om}^{(Q_0)}(\gamma) \;,
\end{equation}
where $\tilde{R}$ is a properly chosen $\gamma$- and $\om$-dependent coefficient
and $f_{\om}^{(Q_0)}(\gamma)$ denotes the unintegrated gluon density in
$\gamma$-space in the $Q_0$-scheme. The latter is directly provided by the
$\e=0$ small-$x$ BFKL equation, possibly of resummed (RGI) type. It is then
clear that, at LL$x$ level, $\tilde{R}$ in Eq.~(\ref{ktAnsatz}) takes the
saddle point value $\tilde{R}\big(\gammaL(\asb(t)/\om),0)$, which therefore
should coincide with the expression~(\ref{R0}) in order to reproduce
Eq.~(\ref{schemeRel}). Furthermore, the NL$x$ expression~(\ref{rk}) can be
reproduced too, by a properly chosen $\ord{\om}$ term in the expression of
$\tilde{R}$; and similarly for further subleading terms in the $\om$-expansion
of $\tilde{R}$.  In other words, Eq.~(\ref{ktAnsatz}) can be made equivalent to
Eq.~(\ref{rk}) at any degree of accuracy in the logarithmic small-$x$ hierarchy,
but has the advantage that the effective anomalous dimension is order by order
dictated by $f_{\om}^{(Q_0)}$, possibly in RGI resummed form, and is therefore
much smoother than its LL$x$ counterpart.

\begin{figure}[htbp]
  \centering
  \includegraphics[width=0.55\textwidth]{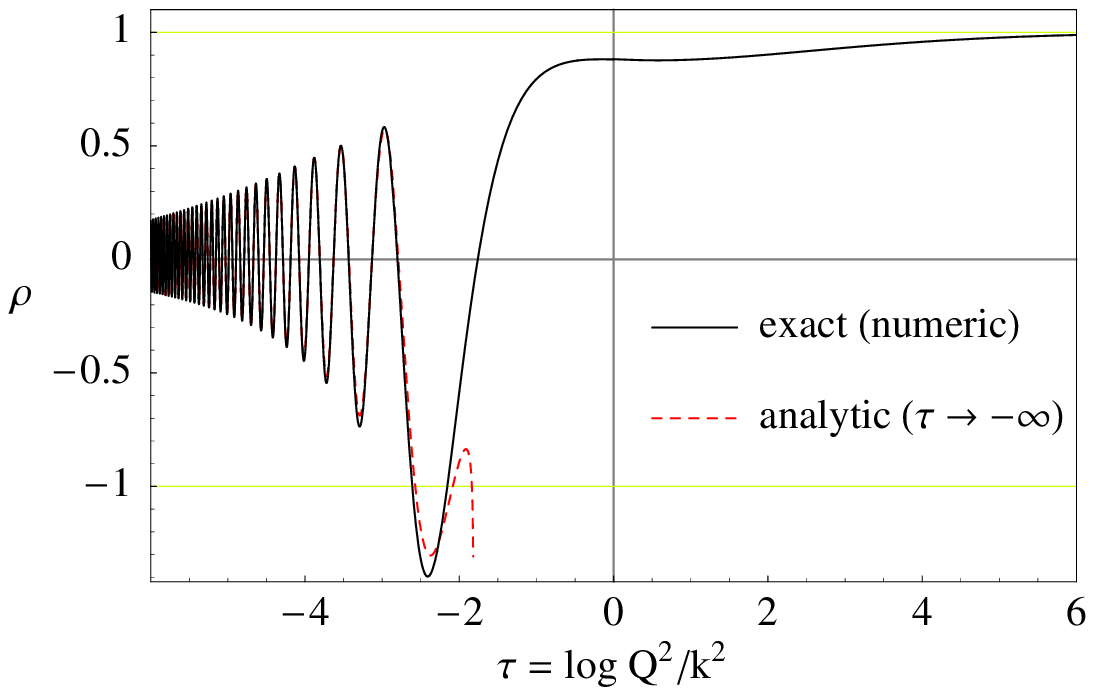}
  \hspace{5mm}
  \includegraphics[width=0.4\textwidth,height=0.35\textwidth]{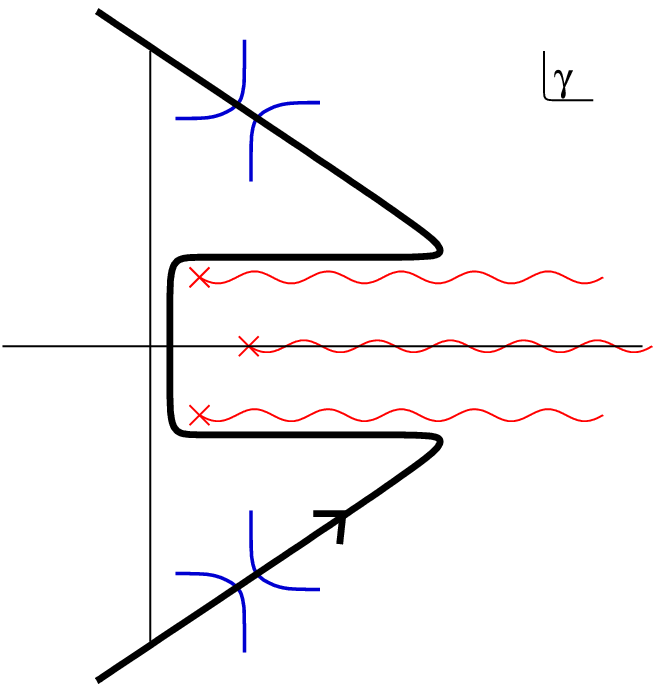}
  \caption{
    {\it (a)} The function $\rho(\tau)$ (solid) and its asymptotic estimates for
    $\tau\to-\infty$ (dashed).
    {\it (b)} Sketch of the fastest convergence contour in the complex
    $\gamma$-plane of the integral in Eq.~(\ref{d:rho}) for $\tau \ll
    -1$, including the cuts and two saddle points that provide the
    dominant contributions to $\rho$ for very negative $\tau$.}
  \label{f:rho}
\end{figure}

By the argument above, we expect the $\om$-expansion of the $\kt$-factorized
scheme-change~(\ref{ktAnsatz}) to be more convergent than~(\ref{rk}) in the
small-$x$ region. This encourages us to implement it at leading level ($\om=0$),
in which we have
\begin{equation}\label{LLconv}
 g_{\om}^{(\MSbar)}(t) = \int_{-\infty}^{+\infty}
 \rho(t-t') f_{\om}^{(Q_0)}(t') \; \dif t' \;,
\end{equation}
where
\begin{equation}\label{d:rho}
 \rho(\tau) = \int_{0^+ + \II}\frac{\dif\gamma}{2\pi\ui} \;
 \esp{\gamma \tau} \frac1{\gamma\,R(\gamma)} \;,
\end{equation}
is pictured in Fig.~\ref{f:rho}.a. For $\tau \geq 0$ it is close to a
$\Theta$-function and for negative $\tau$ it oscillates with a damped
amplitude for larger $|\tau|$ and with increasing frequency. The
difference of $g^{(\MSbar)}$ and $g^{(Q_0)}$ involves a weight
function distributed around $t' \simeq t$ which, convoluted with
$f_{\om}^{(Q_0)}(t')$, includes automatically the RGI resummation
effects of the $Q_0$-scheme.  A word of caution is however needed
about the accuracy of~(\ref{LLconv}) in the finite-$x$ region, where
subleading terms in the $\om$-expansion of $\tilde{R}$ in
Eq.~(\ref{ktAnsatz}) are needed, and are left to future
investigations.

Even if $\Delta(\tau) \equiv \rho(\tau)-\Theta(\tau)$ is in a sense a
small quantity --- because the first two $\tau$-moments of
$\Delta(\tau)$ must vanish --- the numerical evaluation of
(\ref{LLconv}) is delicate because of the large oscillations of $\rho$
in the negative $\tau$ region. For $\tau \ll -1$ the fastest
convergence contour is shown in Fig.~\ref{f:rho}.b, where two saddle
points $\gamma_{\sp}, \gamma_{\sp}^*$ of order $\gamma_{\sp}(\tau)
\simeq 1+\ui\exp\left\{|\tau|+\psi(1)\right\}$ are found. A saddle
point evaluation
\begin{equation}\label{rhosp}
 \rho(\tau)|_{\mathrm{sp}} = \sqrt{\frac{2}{\pi}}
 \Im\left\{\frac{\exp\left[c(\tau_0)+\int_{\tau_0}^\tau \dif \tau' \;
 \gamma_{\sp}(\tau') \right]}{\half\sqrt{\chi_0^{(0)}{}'
 \big(\gamma_{\sp}(\tau)\big)}} \right\} \;,
\end{equation}
provides a good estimate for the contributions from
the diagonal parts of the contour,
while we integrate numerically the remaining part of the contour.  The
result~(\ref{rhosp}) is $\tau_0$-independent, but the constant of
integration $c(\tau_0)$ is determined numerically, e.g., $c(-3) \simeq
-1.18 + \ui\, 2.15$.  The function $\rho(\tau)$ is also computed
entirely numerically for $\tau\gtrsim -7$.

In order to illustrate the difference between small-$x$ gluon
distributions in the $Q_0$ and $\MSbar$ factorisation schemes, we
consider a toy gluon density obtained by inserting a valence-like
inhomogeneous term $f_0$ in the RGI equation
of~\cite{CCSSkernel}, as follows%
\footnote{We adopt a variant of the $\NLLB$ resummation scheme
  introduced in Ref.~\cite{CCSSkernel}, where we perform the
  $\om$-shift also on the higher-twist poles of the NL$x$ eigenvalue;
  we denote it $\NLLBp$. The running coupling, as a function of the
  momentum transfer $q^2$, is cutoff at $q^2=1\GeV^2$.}
\begin{equation}\label{eq:valence-glue}
  f_0(x,t) = A x^{0.5} (1-x)^5 \delta(t - t_0)
  \qquad (e^{t_0} \equiv k_0^2 = 0.55\GeV^2) \;,
\end{equation}
and solving the corresponding evolution for the unintegrated gluon
density $f(x,t) \equiv \G(Q^2,k_0^2;x)$ where $t\equiv \log
Q^2/\mu^2$.  
The normalisation $A$ is set so that the inhomogeneous term
has a momentum sum-rule equal to $1/2$. The solution of the RGI
equation approximately maintains the sum-rule for the full resulting
gluon, though not exactly because of some higher-twist violations.

\begin{figure}[htbp]
  \centering
  \includegraphics[width=0.48\textwidth]{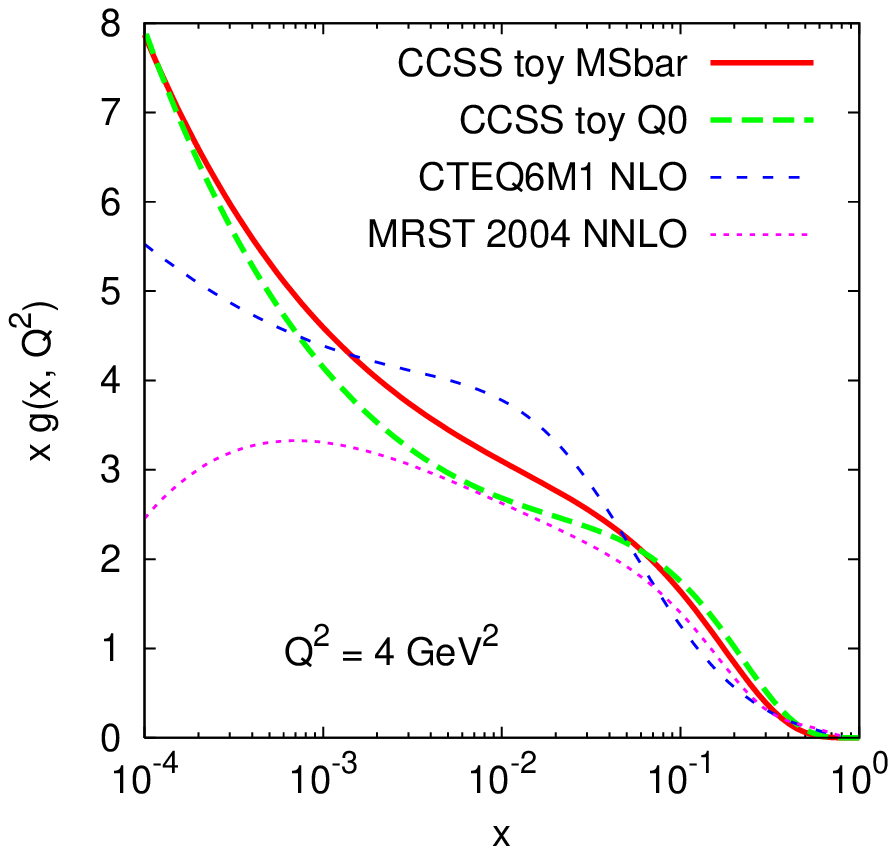}\hfill
  \includegraphics[width=0.49\textwidth]{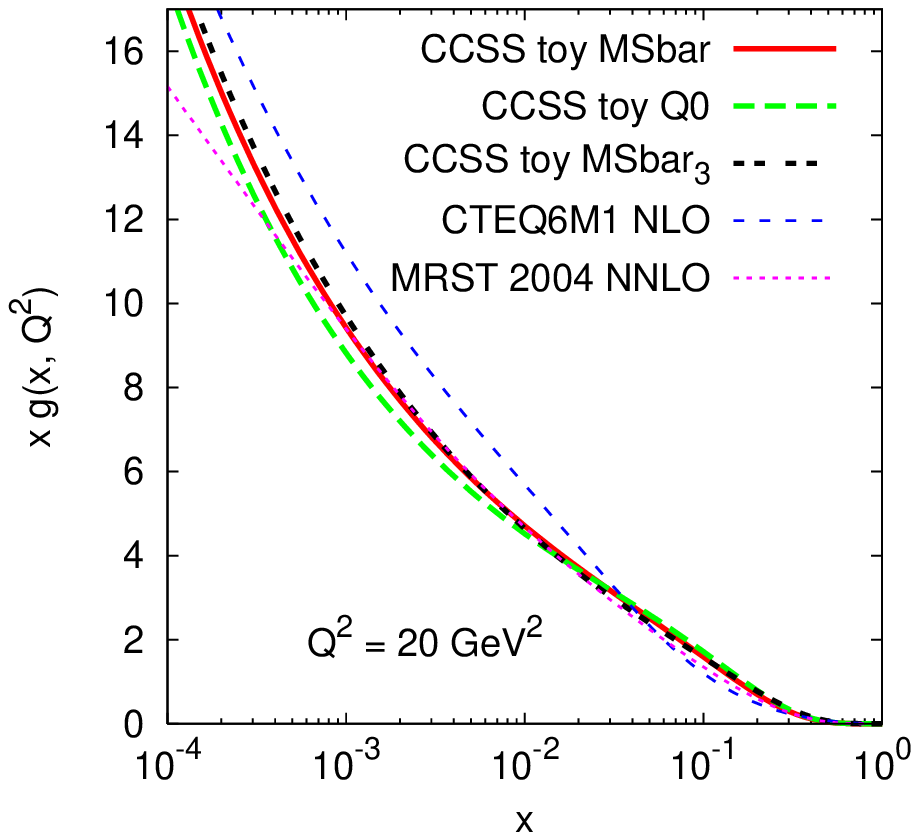}
  \caption{Toy gluon at scales $Q^2 = 4\GeV^2$ and $20\GeV^2$ in the
    $Q_0$ and $\MSbar$ schemes, compared to MRST2004 (NNLO)
    \cite{MRST} and CTEQ6M1 (NLO) \cite{CTEQ}. At the higher $Q^2$
    value we also show the $\MSbar_3$ approximation to the full
    $\MSbar$ evaluation.}
  \label{fig:toy-gluon}
\end{figure}

We then define the integrated densities
\begin{align}
  xg^{(Q_0)}(x,Q^2) &= \int dt'\, \Theta(t - t') \,f(x,t')
 \label{gQ0} \\
  xg^{(\MSbar)}(x,Q^2) &= \int dt' \,\rho(t - t') \,f(x,t')\;,
 \label{gMS}
\end{align}
which are shown in Fig.~\ref{fig:toy-gluon} in comparison to CTEQ
(NLO) and MRST (NNLO) fits for the $\MSbar$ gluon at two scales. Note
that we have not attempted to fine-tune the inhomogeneous gluon term
to get good agreement at large $x$ since in any case we neglect the
quark part of the evolution which is likely to contribute
non-negligibly there.

One observes that the difference between the $Q_0$ and $\MSbar$
schemes is modest compared to that between the CTEQ and MRST fits and,
in particular, there is no tendency for the $\MSbar$ density to go
negative.  This is despite the violently oscillatory nature of $\rho$,
which might have been expected to lead to significant corrections.
This can in part be understood from the approximate form of
$g^{(\MSbar)}$
\begin{equation}\label{eq:MSbar3}
  xg^{(\MSbar_3)}(x,Q^2) \equiv xg^{(Q_0)}(x,Q^2) - \frac{8\zeta(3)}{3}
  \frac{d^3}{dt^3}\big[ xg^{(Q_0)}(x,Q^2) \big] \;,
\end{equation}
which is obtained by expanding the scheme-change in the anomalous
dimension variable $\gamma \sim \dif/\dif t$ to first non-trivial
order. We can see that this approximation is pretty good in the
small-$x$ region for reasonable values of $\as$
($Q^2=20\GeV^2$), corresponding to an effective
$\gamma \simeq 0.25$.%
\footnote{
  At the lower $Q^2$ value we do not show the $\MSbar_3$
  approximation, because in general it breaks down for low $Q^2$.}

\begin{figure}[htbp]
  \centering
  \includegraphics[width=0.48\textwidth]{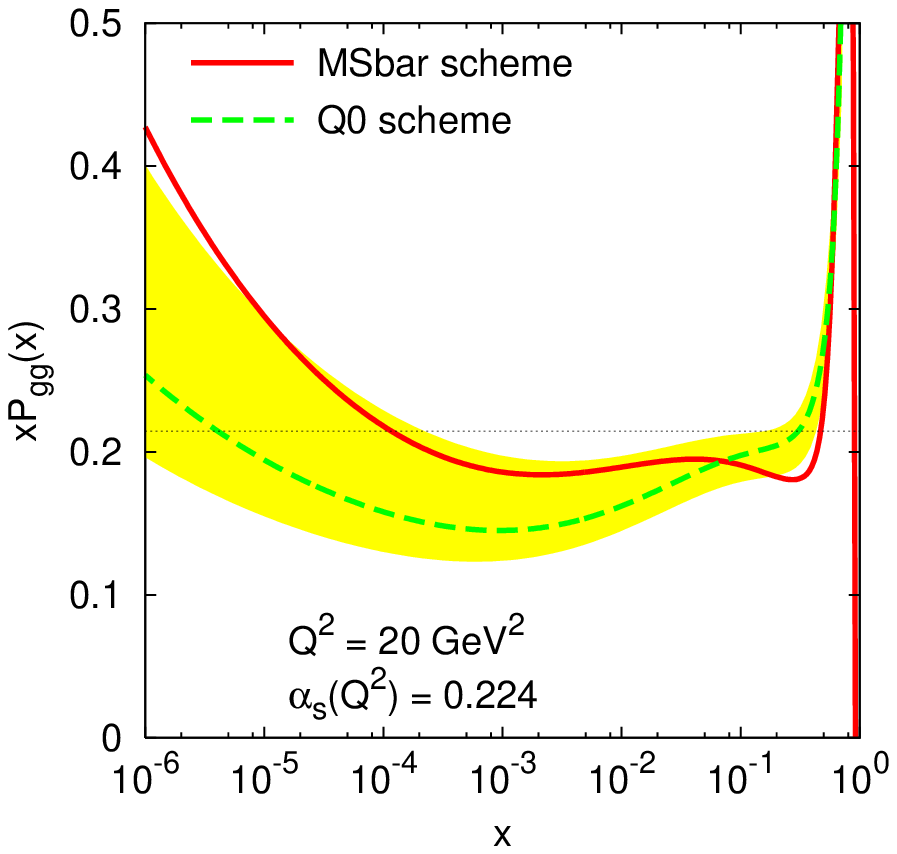}\hfill
  \includegraphics[width=0.48\textwidth]{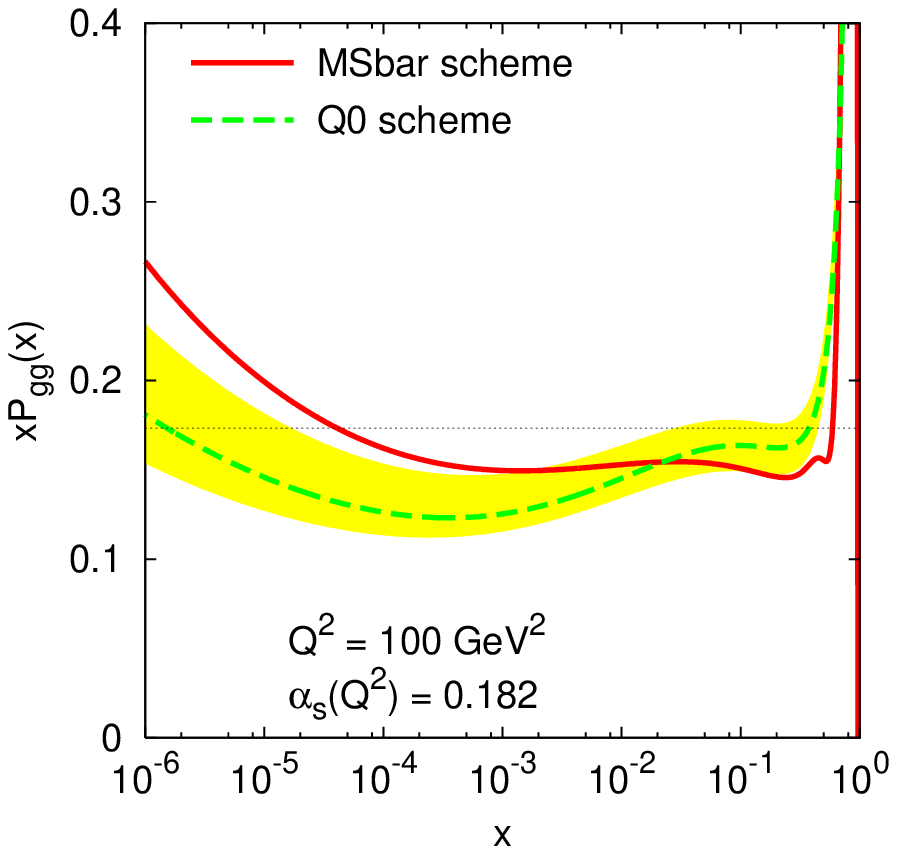}
  \caption{Resummed $P_{gg}$ splitting function in the $Q_0$ and $\MSbar$
    schemes together with the uncertainty band for the $Q_0$ scheme
    that comes from varying the renormalisation scale for
    $\alpha_s$ by a factor $x_\mu$, in the range $1/2 < x_\mu <2$;
    shown for two $Q^2$ values.}
  \label{fig:split}
\end{figure}

We can also use our usual techniques~\cite{deconv} for extracting the
splitting function itself in the $\MSbar$-scheme. The results are
shown in Fig.~\ref{fig:split}, where the $\MSbar$ curve is supposed to
be reliable in the small-$x$ region only ($x \lesssim 10^{-1}$),
because at finite $x$ the $\om$-dependence of the scheme change will
become important.  It appears that the splitting function
is more sensitive to the scheme-change than the density itself. At the
lower $Q^2$ value the difference between the $\MSbar$-scheme and the
$Q_0$-scheme (with $\NLLBp$ resummation) is nearly the same as the
renormalisation scale uncertainty and so might not be considered a
major effect. Recall however that the scale uncertainty is NNL$x$
whereas the factorisation scheme change is a NL$x$ effect, so that
while the renormalisation scale uncertainty decreases quite rapidly as
$Q^2$ is increased (right-hand plot), the effect of the factorisation
scheme change remains non-negligible.

At small $x$, most of the difference between the $\MSbar$ and $Q_0$
splitting functions seems to be due to the $\MSbar$ splitting function
rising as a larger power of $1/x$.
One can check this interpretation by estimating the
factorisation-scheme dependence of the asymptotic $x^{-\om_c}$
behaviour, using the $\MSbar_3$ approximation. We find
\begin{equation}\label{domc}
  \Delta \om_c(t) = \om_c^{\MSbar} - \om_c^{Q_0} \simeq -\frac{\dif
    \om_c(t)}{\dif t} \frac{8\zeta(3)}{3}
  \frac{g'''\big(t,\om_c(t)\big)}{g'\big(t,\om_c(t)\big)} \simeq 0.029\,,
\end{equation}
where the numerical value is given for $Q^2 = 20 \GeV^2$, and is in
rough agreement with what is seen in the left-hand plot of
Fig.~\ref{fig:split}.

\section{Approximate resummed splitting function for the $\boldsymbol\MSbar$ quark}

Let us now discuss the inclusion of  quarks in the resummed, small-$x$
flavour singlet evolution. Resummation effects will be included via
the unintegrated gluon density as discussed before, while the
scheme-change to the $\MSbar$-quark will be an (approximate)
$\kt$-factorised form of the one arising at first non-trivial LL$x$
level.

In order to better specify the scheme change, we first
recall~\cite{CaHa94} that the quark density in a physical $Q_0$-scheme
corresponding to the measuring process $p$ (e.g. $F_2$ or $F_T$ in
DIS) --- which we call $p$-scheme --- is defined, at NL$x$ level, by
$\kt$-factorisation of some $g\to q \bar{q}$ impact factor $\qKernel$,
as follows:
\begin{equation}\label{d:quark}
  q_{\e}^{(p)}(t) \equiv \almu \int \dif^{2+2\e}\kt'\;
  \qKernel(\kt,\kt') \ugd_{\e}(\kt') \;.
\end{equation}
By then working out this convolution in terms of the eigenvalue
function $\qcf_{\e}(\gamma)/\gamma(\gamma+\e)$ of $\qKernel$, one
directly obtains a NL$x$ resummation formula for the $\e$-dependent
$qg$ anomalous dimension in the $p$-scheme:
\begin{equation}\label{d:gammaP}
  \left[ g_{\e}^{(Q_0)}(t) \right]^{-1} \frac{\dif}{\dif t} q_{\e}^{(p)}(t)
  \equiv \gamma_{qg}^{(p)}\big(\as(t),\om;\e\big) \;,
\end{equation}
where, in the $\e = 0$ limit, $\gamma_{qg}^{(p)} = \as(t) \qcf_0(\gammaL)$ has
been obtained in closed form in various cases~\cite{CaHa94,dimensional},
including sometimes the $\e$-dependence.

The $\MSbar$-scheme is then related to the $p$-scheme by the transformation
\begin{equation}\label{pMS}
  q^{(p)} = C_{qq}^{(p)} q^{(\MSbar)} + C_{qg}^{(p)} g^{(\MSbar)} \; ,
\end{equation}
where we set $C_{qq}^{(p)} = 1$ and $b = 0$ at NL$x$ level, so that
$\as(t)=\almu \esp{\e t}$. We thus obtain, by
the definition (\ref{d:gammaP}) and by Eq.~(\ref{schemeRel}),
\begin{align}
  \gamma_{qg}^{(p)}(\as,\om;\e) R(\as,\om;\e)
  &= \left(g_{\e}^{(\MSbar)}\right)^{-1} \frac{\dif}{\dif t} \left[
  C_{qg}^{(p)}(\as,\om;\e) g_{\e}^{(\MSbar)} \right]
  + \gamma_{qg}^{(\MSbar)}(\as,\om)
 \nonumber \\
  &= \left[\gammaL\big(\frac{\asb}{\om}\big) + \e \hat{D} \right]
  C_{qg}^{(p)}(\as,\om;\e) + \gamma_{qg}^{(\MSbar)}(\as,\om) \;,
 \label{d:gammaMS}
\end{align}
where $\hat{D} = \as\partial/\partial\as$ and $\gamma_{qg}^{(\MSbar)}$
is universal and $\e$-independent, so that the process dependence of
$\gamma_{qg}^{(p)}$ is carried by the perturbative, scheme-changing
coefficient $C_{qg}^{(p)}$.

The coefficient $C_{qg}^{(p)}$ in Eq.~(\ref{d:gammaMS}) can be
formally eliminated by promoting $\e$ to be an operator in
$\as$-space, as follows:
\begin{equation}\label{epsOp}
 \e = -\gammaL\big(\frac{\asb}{\om}\big) \hat{D}^{-1} \; ,
\end{equation}
so that the square bracket in Eq.~(\ref{d:gammaMS}) in front of $C_{qg}^{(p)}$
just vanishes (this procedure is rigorously justified in~\cite{dimensional}).
That implies that the l.h.s.\ of Eq.~(\ref{d:gammaMS}) and $\qcf_{\e}(\gamma)$
are to be evaluated at values of $\e \sim \gammaL = \ord{\as/\om}$ which
modifies $\gamma_{qg}^{(\MSbar)}-\gamma_{qg}^{(p)}$ at relative LL$x$ level.
Therefore, even if $\gamma_{qg}^{(\MSbar)}$ is by itself a NL$x$ quantity, the
subtraction of the coefficient part in Eq.~(\ref{d:gammaMS}) affects the scheme
change at relative leading order, contrary to the gluon case of
Eqs.~(\ref{gammaMSbar}) and (\ref{diffAnomDim}).

The replacement (\ref{epsOp}) was used in~\cite{dimensional} in order to get an
``exact'' expression of $\gamma_{qg}^{(\MSbar)}$ at NL$x$ level (that is,
resumming the series of next-to-leading $\om$-singularities
$\sim \as(t) [\as(t)/\om]^n$). The main observation is that, by expanding 
$\qcf_{\e}(\gamma)$ in the $\gamma$ variable around $\gamma=-\e$ with
coefficients $\qcf_n(\e)$, and by converting Eq.~(\ref{d:quark}) to
$\gamma$-space, we obtain
\begin{align}
 \frac{\dif}{\dif t} q^{(p)}_{\e}(t) &=
 \as(t) \int\dif\gamma\; \esp{\gamma t} \qcf_{\e}(\gamma)
 \tilde{g}_{\e}(\gamma,\om)
 \nonumber \\ \label{qDotGamma}
 &= \left( \qRes(\e) + \sum_{n=1}^{\infty} \qcf_n(\e)
 \frac{\dif^n}{\dif t^n}\right) \left( \as(t) R\Big(\frac{\asb}{\om},\e\Big)
 g_{\e}^{(\MSbar)}(t,\om) \right) \;,
\end{align}
where $\tilde{g}_{\e}(\gamma,\om)$ is the Fourier transform of
$g_{\e}^{(\MSbar)}(t,\om)$ and $\qRes(\e) \equiv \qcf_{\e}(-\e)$ turns out to be
the {\em universal} function
\begin{equation}\label{qKernelEps}
 \qRes(\e) = \left[ \frac{T_R}{2\pi} \, \frac{2}{3} \,
 \frac{1+\e}{(1+2\e)(1+\frac{2}{3}\e)} \right]
 \left[ \frac{\esp{\e\psi(1)}\Gamma^2(1+\e)\Gamma(1-\e)}{\Gamma(1+2\e)} \right] 
 \equiv \qResR(\e) \qResT(\e)\; ,
\end{equation}
which we factorise into parts with rational and transcendental coefficients.
Note that the universality of $\qRes(\e)$, which is proportional to the residue
of the characteristic function~$\qcf_{\e}(\gamma)/\gamma(\gamma+\e)$ at the
collinear pole $\gamma = -\e$, is due to the interesting fact that one can
define~\cite{CaHa94} a universal~\cite{dimensional}, off-shell
$g(\kt) \to q(\lt)$ splitting function in the collinear limit
$\kt^2 \sim \lt^2 \ll Q^2$, for any ratio $\kt^2/\lt^2$ of the corresponding
virtualities.

We then realise from Eq.~(\ref{qDotGamma}) that the terms in the r.h.s.\ with at
least one derivative $\dif/\dif t$ are of coefficient type, and vanish by the
replacement~(\ref{epsOp}). The remaining one, proportional to $\qRes(\e)$, is
universal and, by~(\ref{epsOp}), yields the NL$x$ result~\cite{dimensional}
\begin{equation}\label{gqgMSbar}
  \gamma_{qg}^{(\MSbar)}\big(\as(t),\om\big)
  = \as(t) \qResR\bigg( -\gammaL\Big(\frac{\asb(t)}{\om}\Big)
  \frac1{1+\hat{D}} \bigg) \sum_{n=0}^{\infty} \left(
  \gammaL\Big(\frac{\asb(t)}{\om}\Big) \frac1{1+\hat{D}} \right)^n
  R_n\Big(\frac{\asb(t)}{\om}\Big) \;,
\end{equation}
where we have factorised $\as(t)$ by the commutator $[\hat{D},\as]=\as$ and we
have defined the quantity
\begin{equation}\label{Rexpns}
  \qResT(\e) R\Big(\frac{\asb}{\om},\e\Big)
  = \sum_{n=0}^{\infty} (-\e)^n R_n\Big(\frac{\asb}{\om}\Big) \;,
\end{equation}
which has transcendental coefficients and differs from unity by terms of order
$(\gammaL)^3$ or higher, in the $\e \sim \gammaL$
region~\cite{CaHa94,dimensional}.

The complicated expression~(\ref{gqgMSbar}) has been evaluated
iteratively~\cite{CaHa94}, but not resummed in closed form. However, it can be
drastically simplified in the $\kt$-factorised framework by neglecting the
transcendental corrections $\ord{\gamma^3}$, on the ground that the resummed
anomalous dimension is small. In fact, by setting $R=\qResT=1$ and
$\gammaL = \asb/\om$, Eq.~(\ref{gqgMSbar}) reduces to the expression
\begin{equation}\label{borel}
 \gamma_{qg}^{(\MSbar)} \simeq \as(t) \qResR\Big(-\frac{\asb}{\om}
 \frac1{1+\hat{D}} \Big) \cdot 1 = \as(t) \sum_n
 \frac{\qResR_n}{n!} \left(\frac{\asb}{\om}\right)^n \;,
\end{equation}
which is just the Borel transform of
\begin{equation}\label{hRat}
 \qResR(\e) = \sum_n \qResR_n\,(-\e)^n = \frac{T_R}{4\pi}
 \left[\frac1{1+2\e} + \frac13 \, \frac1{1+\frac23 \e} \right] \;.
\end{equation}
Since $\qResR_n = \frac{T_R}{4\pi} [2^n+\frac13(\frac23)^n]$ we obtain
from~(\ref{borel}) what we call the ``rational'' approximation
NL$x|_{\mathrm{rat}}$ (first derived in~\cite{CaHa94})
\begin{equation}\label{GqgRatGamma}
 \left.\gamma_{qg}^{(\MSbar)}\right|_{\mathrm{rat}}
 \simeq \frac{\as(t) T_R}{4\pi} \left(
 \esp{2\frac{\asb}{\om}} + \frac13 \esp{\frac23 \frac{\asb}{\om}} \right) \;.
\end{equation}
This result can be further interpreted in $\kt$-factorised form
\begin{equation}\label{qkt}
  \frac{\dif \qRat}{\dif t} = H^{(\MSbar)}_{\mathrm{rat}} \otimes f \;,
\end{equation}
by using the characteristic function
\begin{equation}\label{Hms}
  \qRes^{(\MSbar)}_{\mathrm{rat}}(\gamma) = \frac{\as T_R}{4\pi} \frac1{\gamma}
  \left( \esp{2\gamma} + \frac13 \esp{\frac23 \gamma} \right) \;,
\end{equation}
which yields the result in Eq.~(\ref{GqgRatGamma}) at the LL saddle point.

\begin{figure}[htbp]
  \centering
  \includegraphics[height=0.49\textwidth]{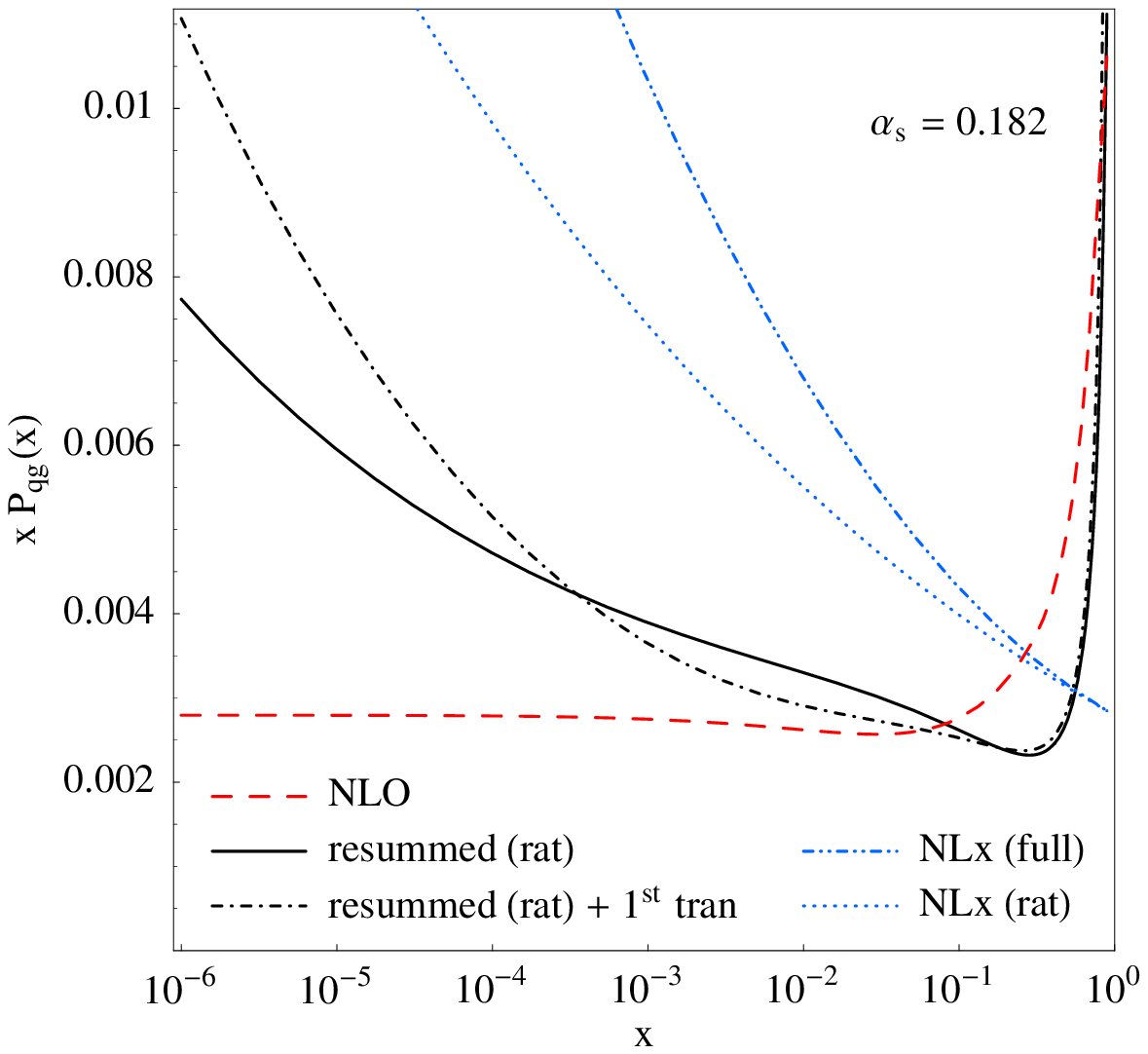}\hfill
  \includegraphics[height=0.494\textwidth]{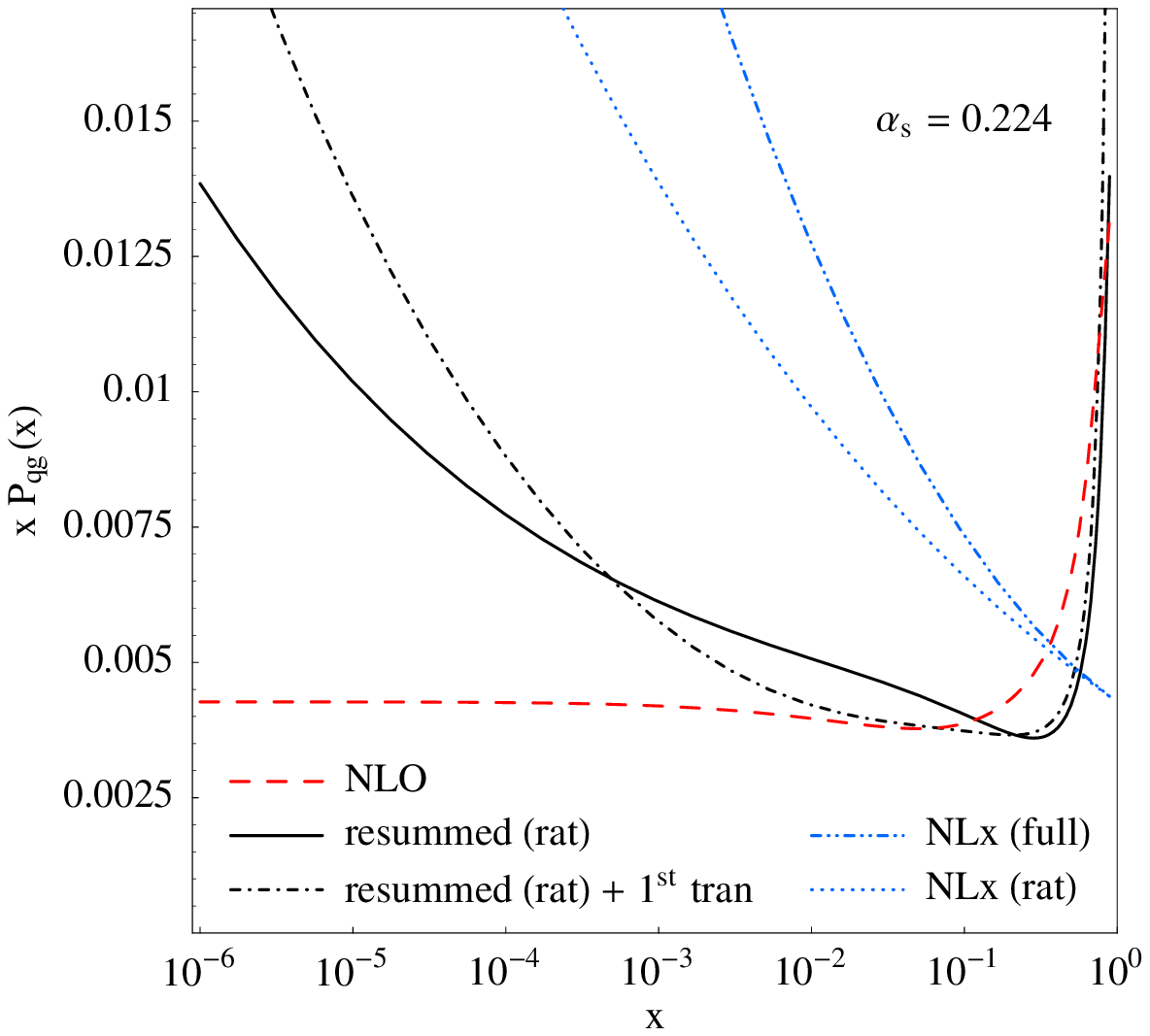}
  \caption{The $\MSbar$ $g \to q$ splitting function for two values of
    $\as$ and in various approximations:
    at two-loop [dashed], the RGI resummed rational one in Eq.~(\ref{PqgRat})
    [solid] and with the addition of the first transcendental correction
    [dash-dotted], the rational NL$x$ approximation [dotted] and the complete
    NL$x$ one [dash-dot-dotted]}
  \label{f:toyPqg}
\end{figure}

Finally, since the exponentials in~(\ref{Hms}) generate translations
in $t$, our rough estimate of the resummed $P_{qg}$ is provided by the
simple formula
\begin{subequations}\label{PqgRat}
\begin{align}
  \frac{\dif}{\dif t} \qRat(t,x) &= \frac{\as(t) T_R}{4\pi}
  \left[g\big(t+2,x\big) + \frac13 g\big(t+\frac23,x \big) \right]
 \label{qDot} \\
 &= \int_x^1 \frac{\dif z}{z} \; P_{qg}^{(\MSbar)}\big(\as(t),z\big)
 g\big(t,\frac{x}{z} \big) \;.
\end{align}
\end{subequations}
By replacing in Eq.~(\ref{qDot}) the resummed gluon
density~\cite{CCSSkernel} and by performing the necessary
deconvolution~\cite{deconv} we obtain the results in
Fig.~\ref{f:toyPqg}, compared to NLO and NL$x$~\cite{CaHa94} results.

A proper comparison of the curves in Fig.~\ref{f:toyPqg} can be made only in the
small-$x$ region $x \lesssim 10^{-1}$ because Eqs.~(\ref{gqgMSbar}) and
(\ref{PqgRat}) do not include the finite-$x$ perturbative terms.
We then notice that small-$x$ resummation effects in $\qRat$ are sizeable
even around $x \sim 10^{-3}$ and somewhat larger than the gluonic
ones. They are anyway much smaller than the corresponding ones of the
NL$x$ result, thus showing that the resummed anomalous dimension
variable is pretty small, as already noticed in the gluon case. We can
also check how good the ``rational'' resummation is, by calculating
the effect of the first $\ord{\gamma^3}$ transcendental correction,
which is also shown in Fig.~\ref{f:toyPqg}. It appears that the
difference is indeed not large and anyway much smaller than the
difference between the results of NL$x|_{\mathrm{rat}}$ in
Eq.~(\ref{GqgRatGamma}) and NL$x$~\cite{CaHa94}, both shown in
Fig.~\ref{f:toyPqg}.

\vspace{5mm}

To sum up, we have proposed here a $\kt$-factorised form of the $Q_0
\to \MSbar$ scheme-change (Eqs.~(\ref{ktAnsatz}) and (\ref{qkt}))
which allows a convergent leading log hierarchy, because of the
smoothness of the resummed anomalous dimension. Applying the leading
scheme change to the gluon density and --- in an approximate way ---
to the quark density we have provided predictions for the $gg$ and
$qg$ splitting functions in the $\MSbar$-scheme and for the
corresponding densities.

We find that the gluon density itself is rather insensitive to the
scheme change, while its splitting function is somewhat sensitive.
Resummation effects for the $\MSbar$ quark are more important, but
anyway much smaller than those at NL$x$ level. We are thus confident
that the scheme change can be calculated in a reliable way in a fully
resummed approach as well.

\section*{Acknowledgments}

We thank John Collins for a question about positivity of the gluon in
the $\MSbar$ scheme which motivated us, in part, to push forwards with
these investigations.
This research has been partially supported by the Polish Committee for
Scientific Research, KBN Grant No. 1 P03B 028 28, by the U. S.
Department of Energy, Contract No. DE-AC02-98CH10886 and by MIUR
(Italy).


\end{document}